# Fracture toughness of leaves: Overview and observations

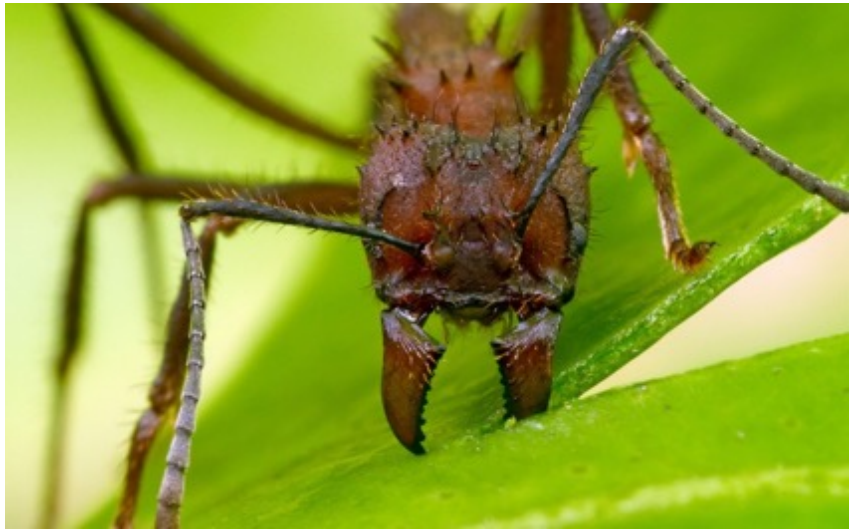

**By: Mehrashk Meidani**

cite as: arXiv:1601.00979 [q-bio.TO]

# Table of Contents

# Fracture toughness of leaves: Overview and observations

**By Mehrashk Meidani**[1]

Department of Civil & Environmental Engineering

University of Massachusetts Amherst

## Abstract

One might ask why is it important to know the mechanism of fracture in leaves when Mother Nature is doing her job perfectly. I could list the following reasons to address that question:

- Leaves are natural composite structures, during millions of years of evolution, they have adapted themselves to their surrounding environment and their design is optimized, one can apply the knowledge gained from studying the fracture mechanism of leaves to the development of new composite materials.

- Other soft tissues like skin and blood vessel have similar structure at some scales and may possess the same fracture mechanism. The gained knowledge can also be applied to these materials.

- Global need for food is skyrocketing. There are few countries, including the United States, that have all the potentials (i.e. water, soil, sunlight, and manpower) to play a major role in the future world food supplying market. If we can increase the output of our farms and forests, by means of protecting them against herbivores [Beck 1965], pathogens [Campbell et al. 1980], and other physical damages, our share of the future market will be higher. It will also enforce our national food security because we will not be dependent on food import. We do not yet know how much of our farms and forests output can be saved if we can genetically design tougher materials, but the whole idea does worth to be studied.

[1] 28 Marston Hall, 131 Natural Resources Road, University of Massachusetts Amherst, Amherst, MA 01003. Email: mmeidani@umass.edu

# 1   Introduction

Leaves have very complex and heterogeneous structure. For each plant, the leaf structure is optimized through millions of years of evolution to get the most of their surrounding environment while surviving harsh conditions including physical and chemical attacks, parasites and foragers.

When studying the fracture mechanism in unconventional materials with complex structure such as leaves, we encounter the following questions: 1- Which fracture mode is dominant?; is there a combined-mode fracture mechanism that governs the fracture? 2- How does the crack grow at vein interfaces, 3- Should cohesive zone be considered in the fracture analyses?

The toughness of a leaf is a measure of energy and time required to cut a part of it. Tougher leaves are less likely to be attacked by foragers. For example leaf-cutting ants give up cutting a leaf piece from the lamina if it takes too long, in order to decrease the risk of being hunted by other predators. The different toughness values in several structural elements of a leaf makes it possible for the leaf to survive attacks and the whole plant will be able to receive partial functionality from the damaged leaves. For instance, the Mason bees try avoiding major veins to minimize the time cutting a certain area of the leaf by following a circular cutting path and (Figure 1). This will leave the leaf alive and useful for the plant.

The cutting behaviors of the aforementioned two insect species are relevant to the purpose of this study. The leafcutter ants (belonging to the two genera Atta and Acromyrmex) cut several portions of the plant leaves and take them to their nest to provide nutritional substrate for their fungal farms (Figure 1). The interesting point in those ants' cutting behavior is that they do not follow any certain pattern in cutting the leaves, except avoiding the very thick and strong primary vein which is either left alone or cut by stronger ants. The same behavior can be seen in the cutting pattern of Mason bees (Megachilidae); they also do not cut across the primary vein. This could be a good clue about fracture toughness of leaves across their cross section where the size and strength of the veins change by several orders of magnitude [Sach et al. 2012]. To develop an appropriate testing method for obtaining the fracture toughness of leaves, the average fracture toughness should be measured and not on a localized region. It may be necessary to avoid cutting through the primary vein, just as those natural foragers do. The mathematical rationale behind this statement is provided in the next section where venation patterns in leaves are discussed in details.

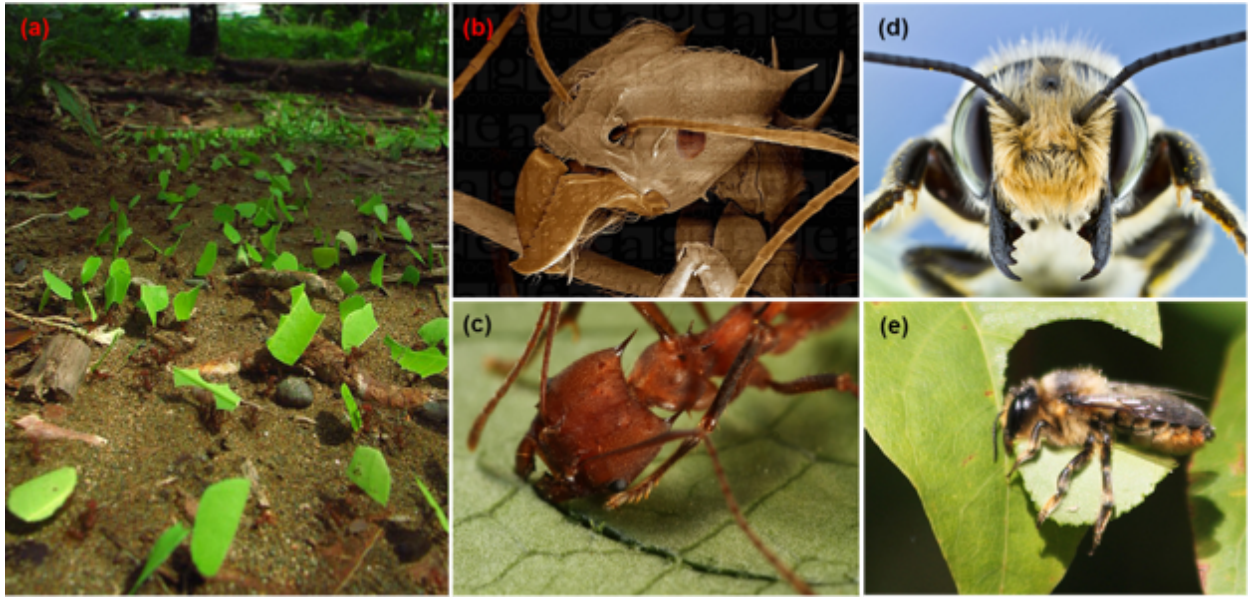

Figure 1. (a) Leafcutter ants transporting leaves [source: Wikimedia, released under the GNU Free Documentation License]; (b) SEM of a leafcutter ant Atta sp., magnification x200 [Photographer: Last Refuge, Collection: Robert Harding Picture Library]; (c) Ant cutting across the leaf, passing through several sized vein [Original source unknown]; (d) Micrograph of a Megachilidae bee head and jaws, aka leafcutter or mason bee [Photo by Phil Sharp].

Conventionally three modes of fracture are defined for materials (Figure 2). Most of the natural fractures in leaves occur in mode III (tearing mode or scissoring mode). However, there are few cases that Mode I fracture (opening mode) may be applied, e.g. when an herbivore animal tries to pull out the grass or leaf from its root or main branch. The fracture toughness of the leaves at these two modes should be obtained from two different types of tests, just like for other types of materials. It is very unlikely that Mode II (sliding mode) fracture occurs in leaves as they are generally very thin and will buckle under any Mode II loading.

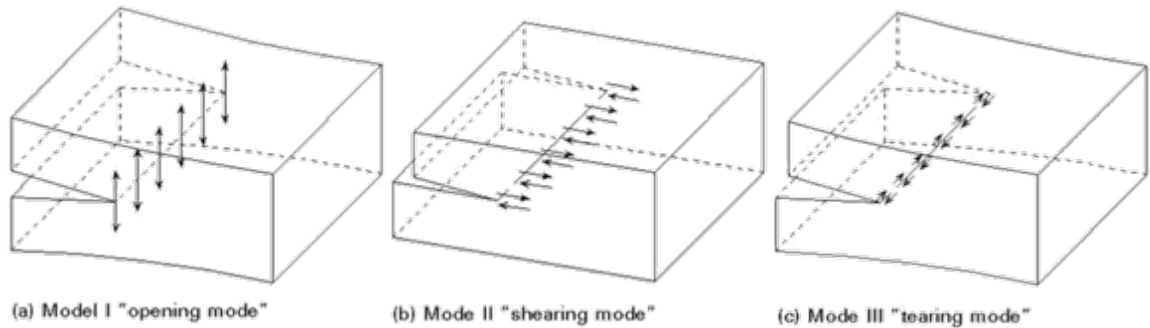

**Figure 2. Three conventional fracture modes.**

Since leaves are complex structural systems, we need to study their components and properties in order to have a better understanding of what occurs in their fracture process.

## 2  Parameters that affect fracture toughness of leaves

Leaf structure is the most important parameter that affects its macro-scale fracture toughness. Biotic and abiotic parameters both affect the leaf strength against fracture. Ambient temperature and humidity are among the abiotic parameters while chemical components of the leaf, e.g. the protein to fiber ratio [Choong et al. 1992], are biotic ones. The fracture toughness of the leaf may vary seasonally as its chemical components change. Also, age of the leaf can affect its fracture toughness [Sach et al. 2012]. Hence, it is crucial to specify leaf conditions, including all aforementioned parameters, at the time of test in order to define a standard method for evaluating the fracture toughness for the leaves.

## 3  Leaf structure

Leaf structure has a major impact on its macro-scale mechanical behavior. Veins constitute the base structure of the leaves with the green membrane material fill in the space between them. There are several characterization methods to categorize leaves according to their venation (Figure 3). It is not practical to investigate fracture toughness mechanism for all of the shown venation patterns here. In this study, cross-venulate and reticulate patterns are studied. Leaves with these venation patterns have more complex composite structure, with structural elements of different sizes.

The venation of the leaf is also a function of its age. In a recent study by Sack et al. (2012), the scaling of leaf venation architecture was studied and explained the global ecological pattern observed in growing leaves of the same species. As can be seen from Figure 5, four levels of veins with different diameters appear in the venation system of 27 species of dicotyledonous plants. Some leaves can have as much as five levels of venation (Figure 4).

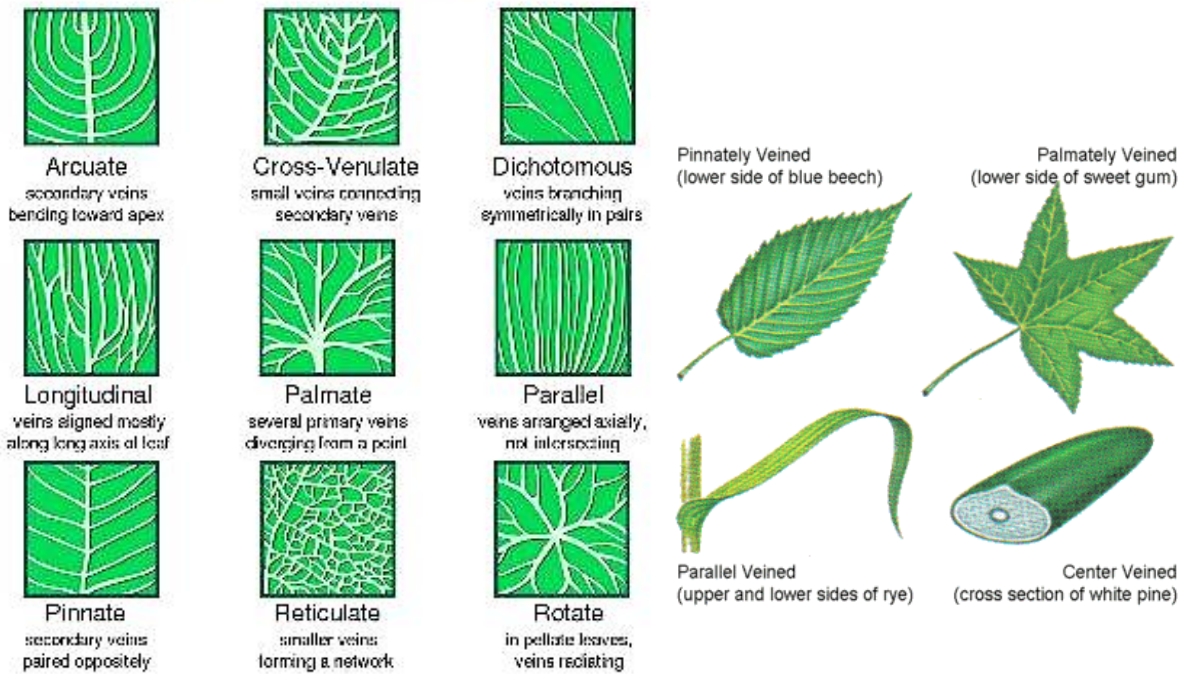

**Figure 3. (Left) different venation patterns in plants leaves [http://science.jrank.org]; (right) Showing four leaf samples with different venations [http://www.robinsonlibrary.com/science].**

Referring to the results of Sack et al. (2012) in Figure 5, the diameter of veins almost halves as it goes to the next level (primary to secondary, secondary to tertiary, etc.). The area which they cover on the surface of a leaf is inversely proportional to the vein diameter. Quaternary veins cover most of the leaf area whilst having the smallest diameter. If we consider the cross section of a leaf, apart from the primary vein, which is obviously larger and stronger than the other parts of the leaf, secondary and tertiary veins almost occupy equal areas, and quaternary and lower order veins cover equal amount of area similar to the membrane (multiply Figure 5b and 4d).

Again, as was suggested earlier from the cutting behavior of the insects, it seems that the fracture toughness of leaves is controlled by their venation system and membrane, put aside the major vein. For a realistic measurement of the fracture toughness in leaves, it is suggested that we do not extend the fracture across the major vein. Two conventional facture toughness evaluation methods are described in the following sections.

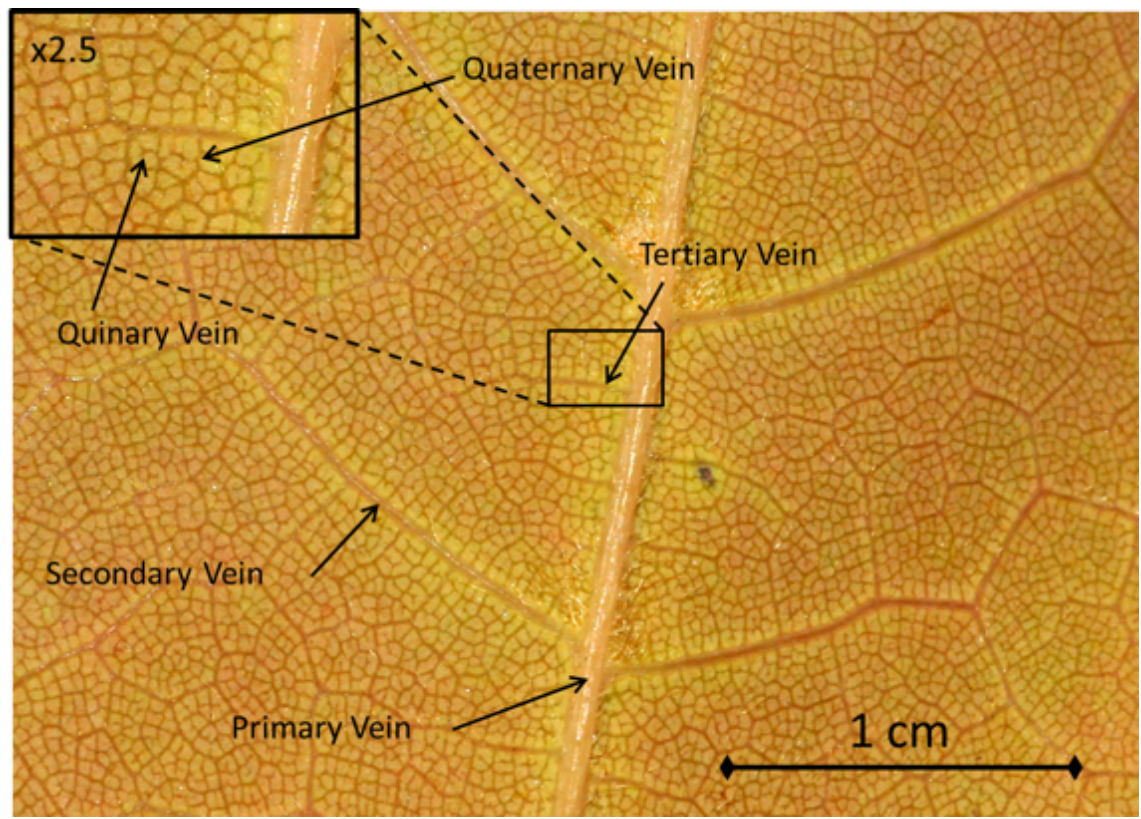

**Figure 4. Five different levels of veins indicated on a Maple tree leaf [Photo by Mehrashk Meidani].**

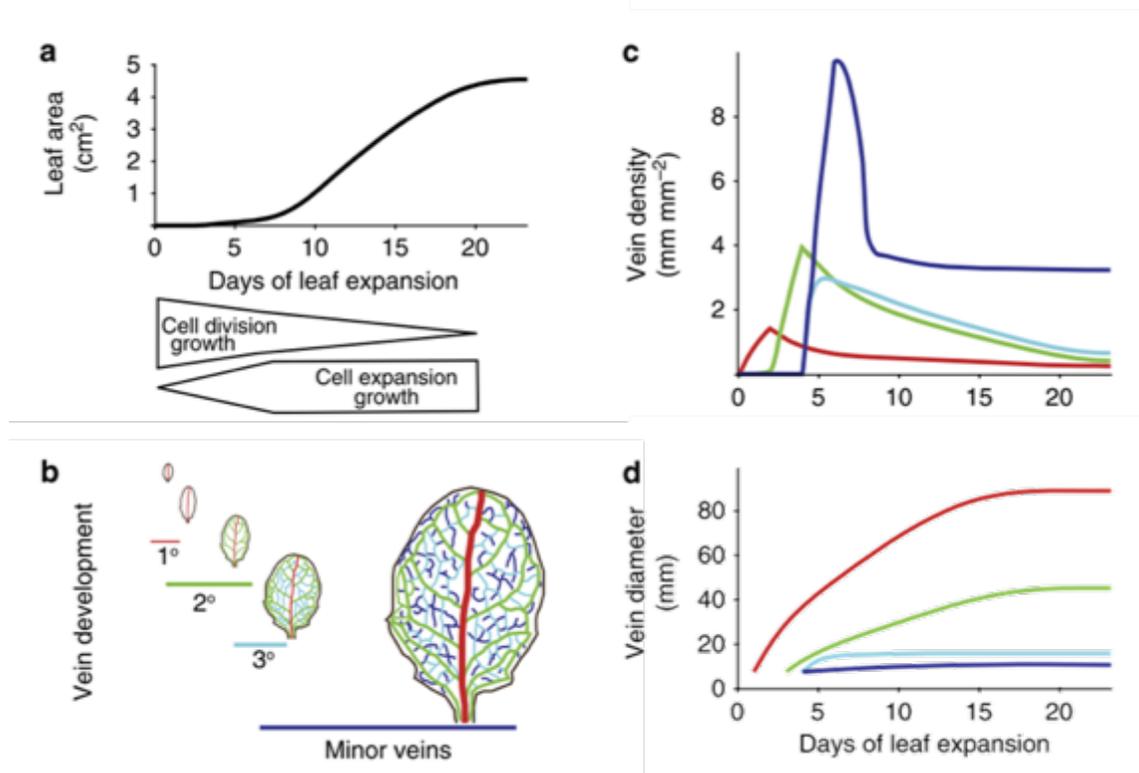

**Figure 5.** Synthetic model for leaf expansion and vein development for Arabidopsis leaves and 27 other dicotyledonous species. (a) expansion of leaf area vs. time; (b) vein orders form in overlapping sequence, The 1° and 2° veins are formed during the slow phase, the 3° veins next, and the minor veins principally during the rapid phase; (c) vein densities of each order peak as procambium forms, then decline as leaf expansion pushes veins apart, but the minor vein density stabilizes as vein initiation is maintained during leaf expansion; (d) The 1° and 2° veins have prolonged diameter growth, whereas 3° veins and minor veins rapidly attain maximum diameter [Sack et al. 2012].

## 4 Common leaf fracture testing methods and devices

### 4.1 Punch-and-die method

In this method, a punch-and-die device is used to evaluate the fracture toughness of a leaf (Figure 6). In the sampling method, the thicker central vein of the leaf is not included. The parts closer to blade edge are avoided too. This specimen could represent an average thickness for the leaf, but overlooking the strength of the central vein. In this test, the fracture is developed under mode III conditions. Force applied to the punch and its axial displacement are plotted and the total energy required to complete the process is calculated as the fracture toughness of the leaf sample.

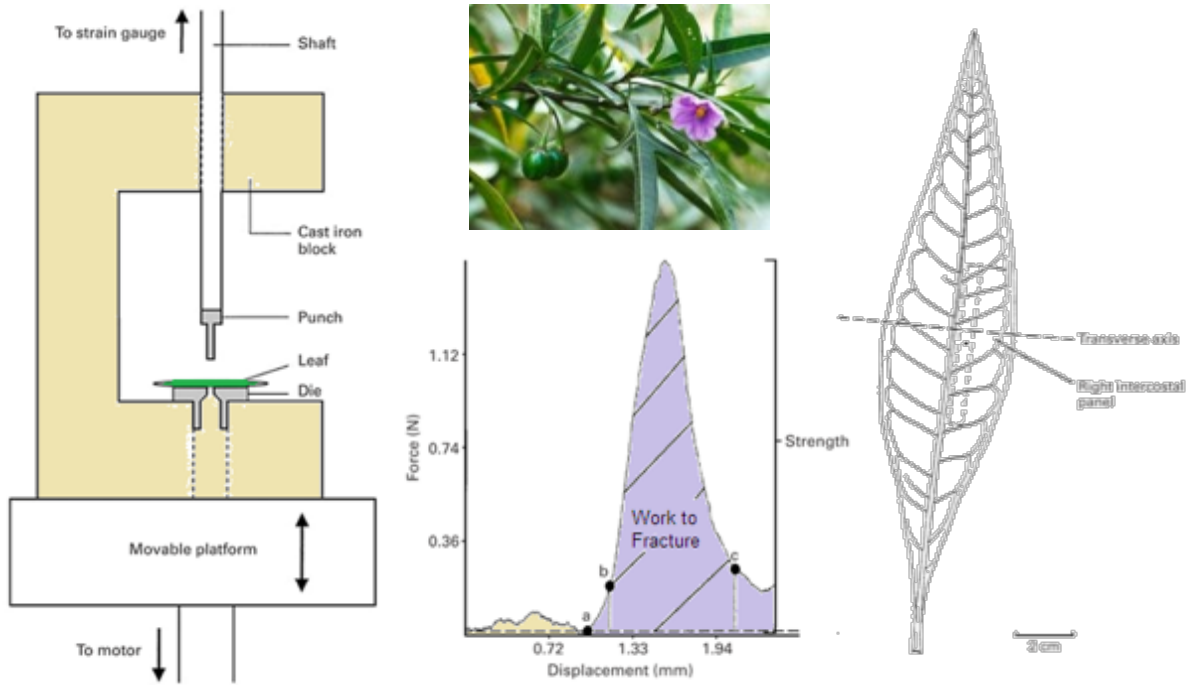

**Figure 6. (Left) The setup of the Chatillon Universal Tester for a punch-and-die test; (mid-top) Solanum laciniatum plant (known as Kangaroo apple); (mid-bottom) a typical force-displacement curve from a punch test on Solanum laciniatum leaf specimen; (right) selecting the test specimen from the leaf.**

## 4.2 Shear-tear method

In this method, the leaf is cut with a razor blade. The cutting force and displacement of the blade is recorded. The fracture toughness of the leaf sample is calculated from the required work to complete the fracture across the specimen. In this test setup, major vein is also included in the crack (cut) path, but not considered in the calculations. It can be seen from Figure 7 that the force required to cut through the midrib is almost one order of magnitude larger than that for the leaf's lamina.

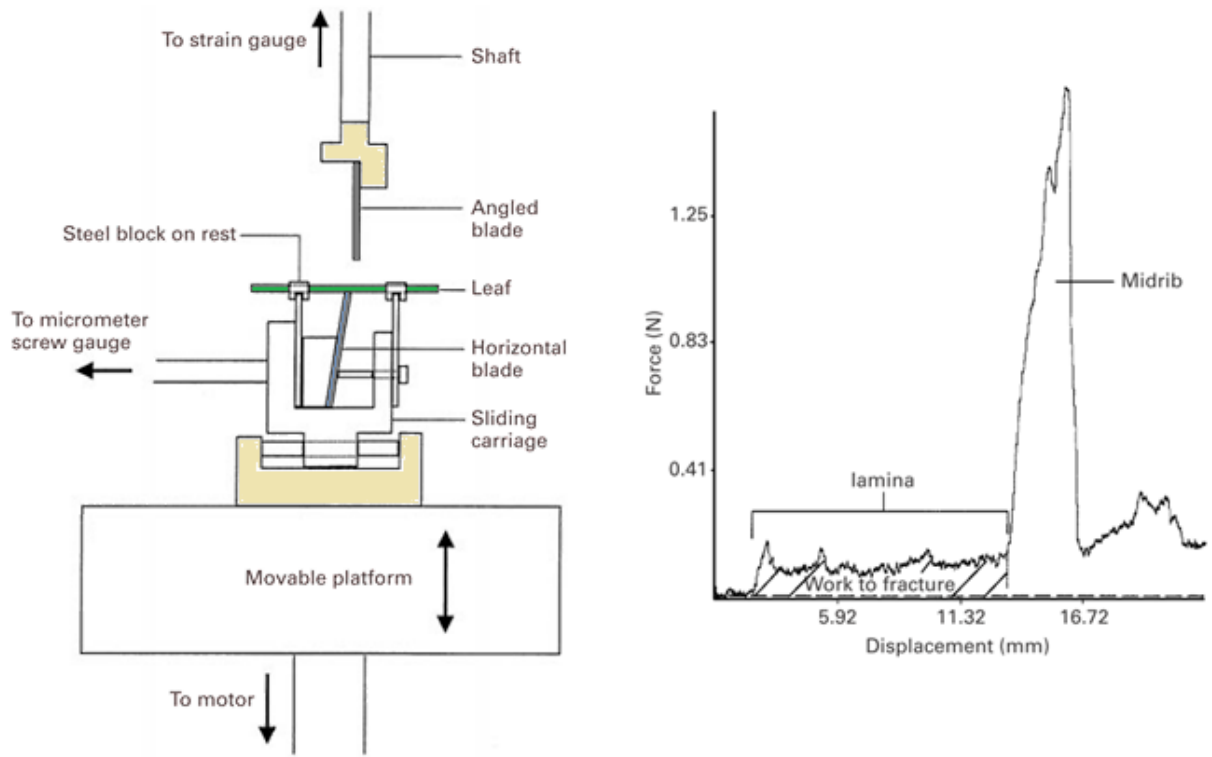

**Figure 7. (Left) the setup of the Chatillon force tester for a shearing test; (right) a force displacement test curve from a shearing test on a Solanum laciniatum leaf specimen [Aranwela et al. 1999].**

## 5  Observations from simple fracture experiments

The foregoing methods used to evaluate the fracture toughness of the leaves have a presumed crack path. In real loading conditions, the crack will follow a path that requires minimum energy. Like any other composite structure, the crack growth path will depend on many parameters including the interface stiffness on different materials. In order to visualize what really occurs during a mode I fracture process in a leaf, the author performed several simple fracture experiments on similar leaf specimens and observed the crack growth under a microscope (Figure 8).

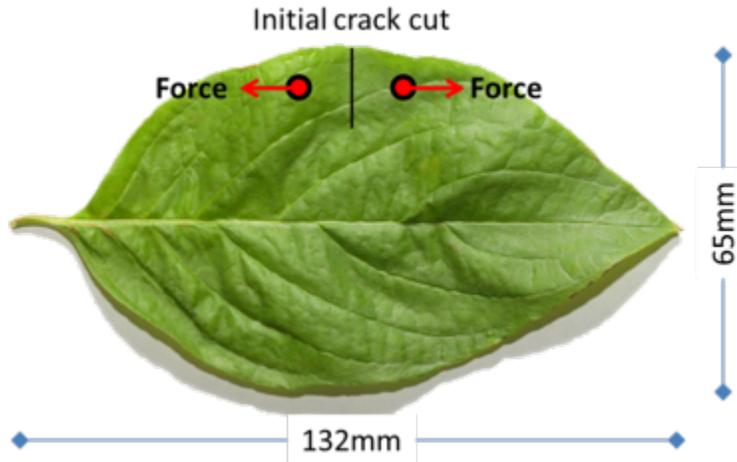

**Figure 8. One of the leaf samples used in fracture observation tests.**

When studying the fracture mechanism in the leaves, we are dealing with highly heterogeneous structures. During crack growth in such a medium, it impinges interfaces joining different materials (i.e. intercoastal material and venation structure). Crack growth may arrest at the interface or may advance in the new material but not follow its previous path (Figure 9). He and Hutchinson (1989a, b) studied this problem in their work "crack deflection at an interface between dissimilar elastic materials" for elastic and isotropic materials on either side of the interface. However, materials in the leaf are neither elastic nor isotropic, and we can just have some insight into the fracture propagation mechanism based on oversimplified assumption.

The problem of crack growth becomes more complicated when the crack impinges the interface with an oblique angle; does the crack tend to deflect along the interface or penetrate into the other material? This peculiarity arises from the formation of singular stress fields for an oblique crack terminating at an interface. In leaves, since they have a non-orthogonal vein structure, it is inevitable for a crack to impinge some of the interfaces with a non-normal angle. So it will most likely follow the interface. Then it is important, for measurement purposes, to kink the crack out of the interface and force it to grow in the stiffer media, otherwise it will continue growing along the vein till it reaches leaf blade or a larger vein.

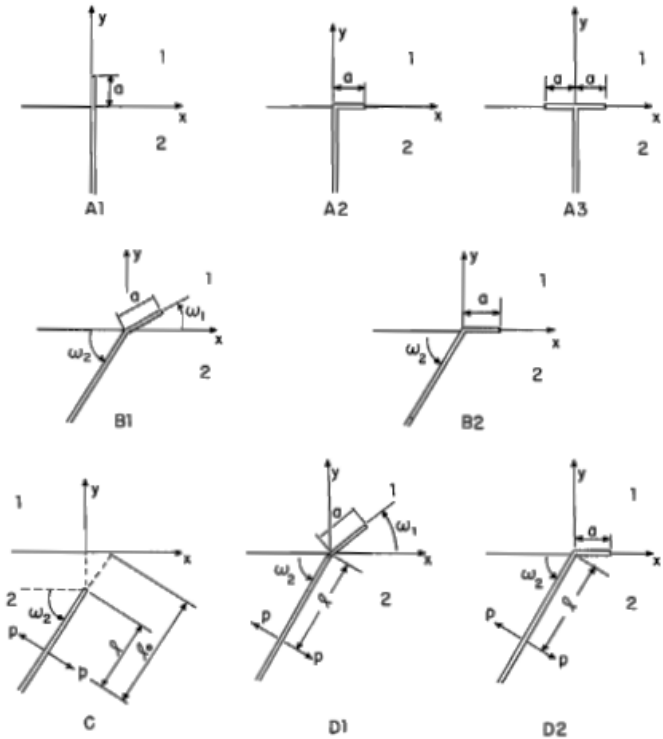

**Figure 9. An existing crack in material 2 impinges the interface with material 1. Several possible crack growth scenarios including crack arrest and deflection are shown in these figures [He and Hutchinson 1989].**

Loading rate has also an important effect of the crack growth path in the leaves; the leaf elements are made of non-linear viscoelastic-viscoplastic materials. If loading rate is very fast, the materials will act stiffer and the crack shape will be different when loading rate is relatively smaller. Based on the microscopic observations (Figure 10), with a slow loading rate, the crack has more tendency to propagate along the interfaces and not cut across the veins. For faster loading rates, the crack will cut across the smaller veins but becomes asymptotic to the major midrib.

## 5.1 Macroscopic observations

To study the effect of loading rate on the crack path in leaves, a leaf is subjected to mode I loading and sequential images with 1.0 second interval are recorded during crack growth (Figure 10). As can be seen from this figure, when crack tip propagating in the membrane material reaches a strong vein, it starts to propagate along their interface. The connection between the membrane and the vein is very complex, consisting of several fibers and ligaments.

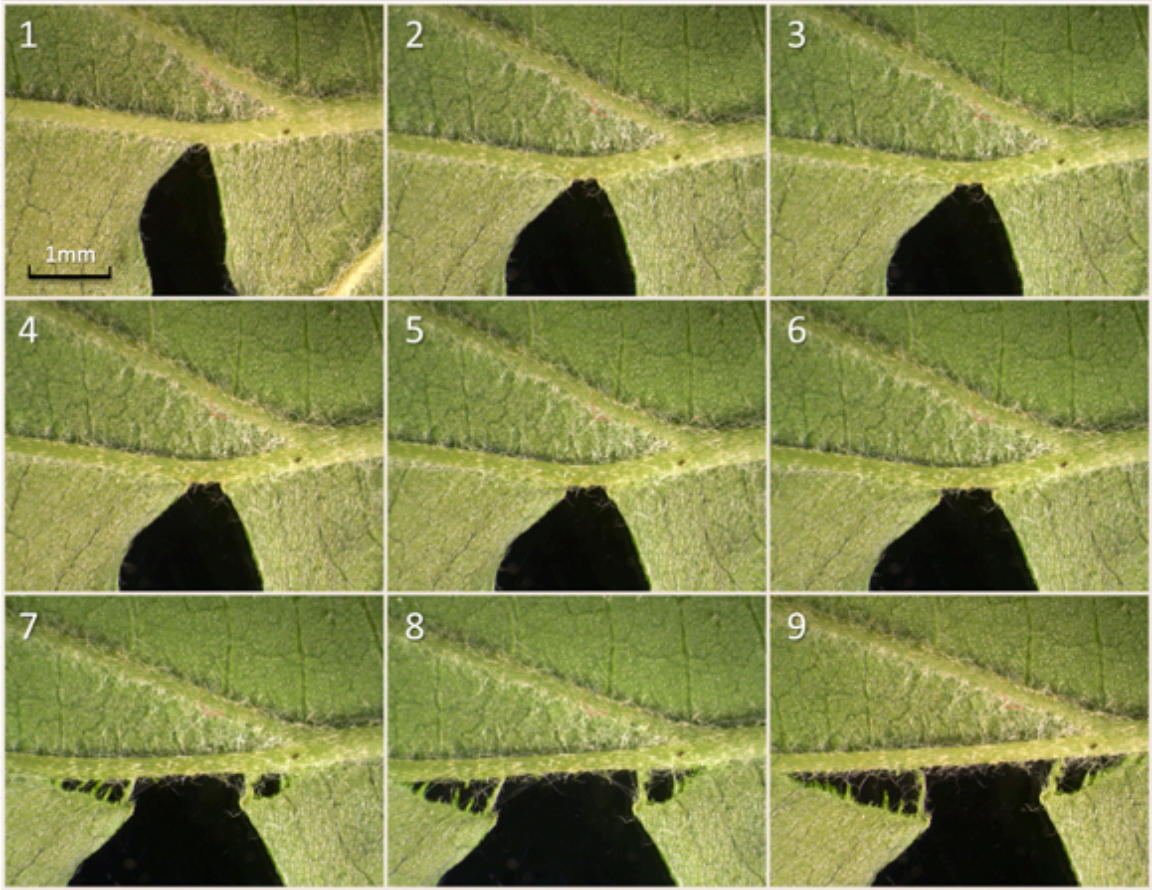

**Figure 10.** Sequential stages of fracture growth in a leaf when crack tip impinges the major vein [Images by Mehrashk Meidani, SEM Center at USC, Columbia].

## 5.2 Microscopic observations

Another similar leaf specimen is put under the microscope with 10x magnification and a relatively higher loading rate is applied to it. The crack growth pattern is different in this case; the crack cuts across the secondary veins and propagates until it reaches the midrib (Figure 11. An important structural element that helps the leaf to keep its detached part alive even after fracture is the group of fibers inside the veins. A closer look at the fractured area on the secondary rib is shown in Figure 12. Despite large strains at the rupture zone, the fibers are still intact and chemical components and water can pass through them to keep the detached part of the leaf alive. These fibrous elements are capable of withstanding very high axial strains. When look at with higher magnification, we can see the spiral structure of the fibers (Figure 13). That is the main reason that they can withstand such high axial strains.

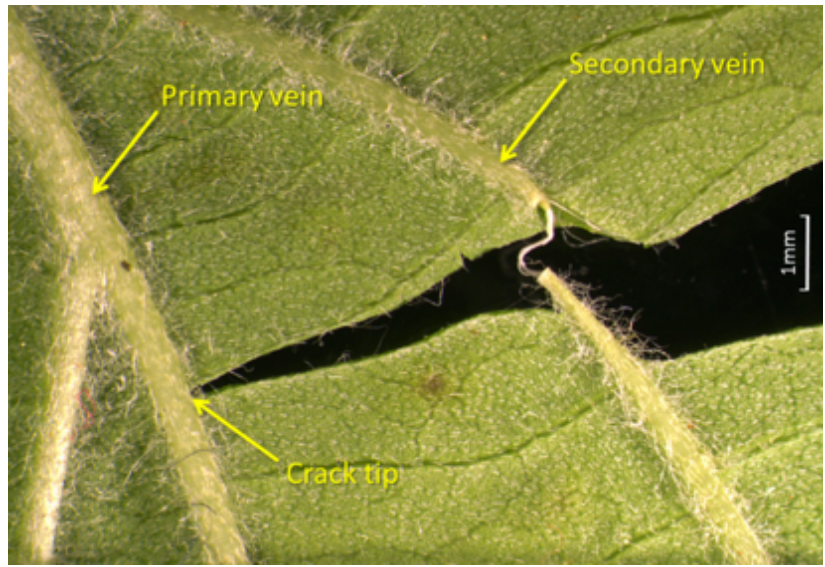

Figure 11. Microscopic image of the crack path across a secondary vein. [Images by Mehrashk Meidani, SEM Center at USC, Columbia].

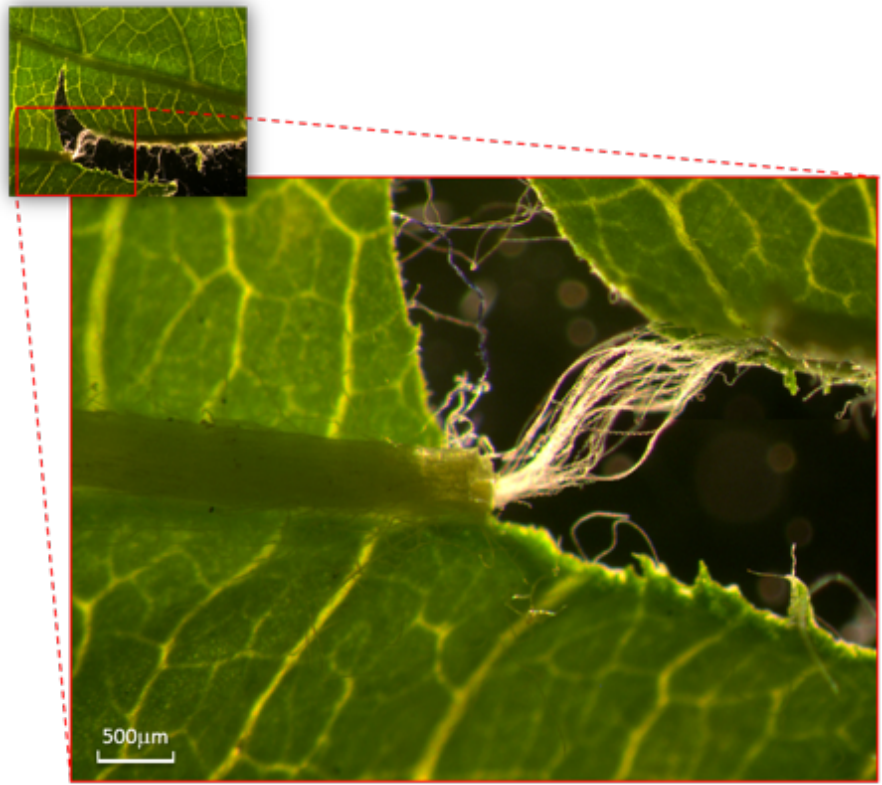

Figure 12. Crack tip cuts through a major vein and continues growing in the intercoastal material. Fibrous material are pulled out of the vein due to the large imposed strains [Image by Mehrashk Meidani, SEM Center at USC, Columbia].

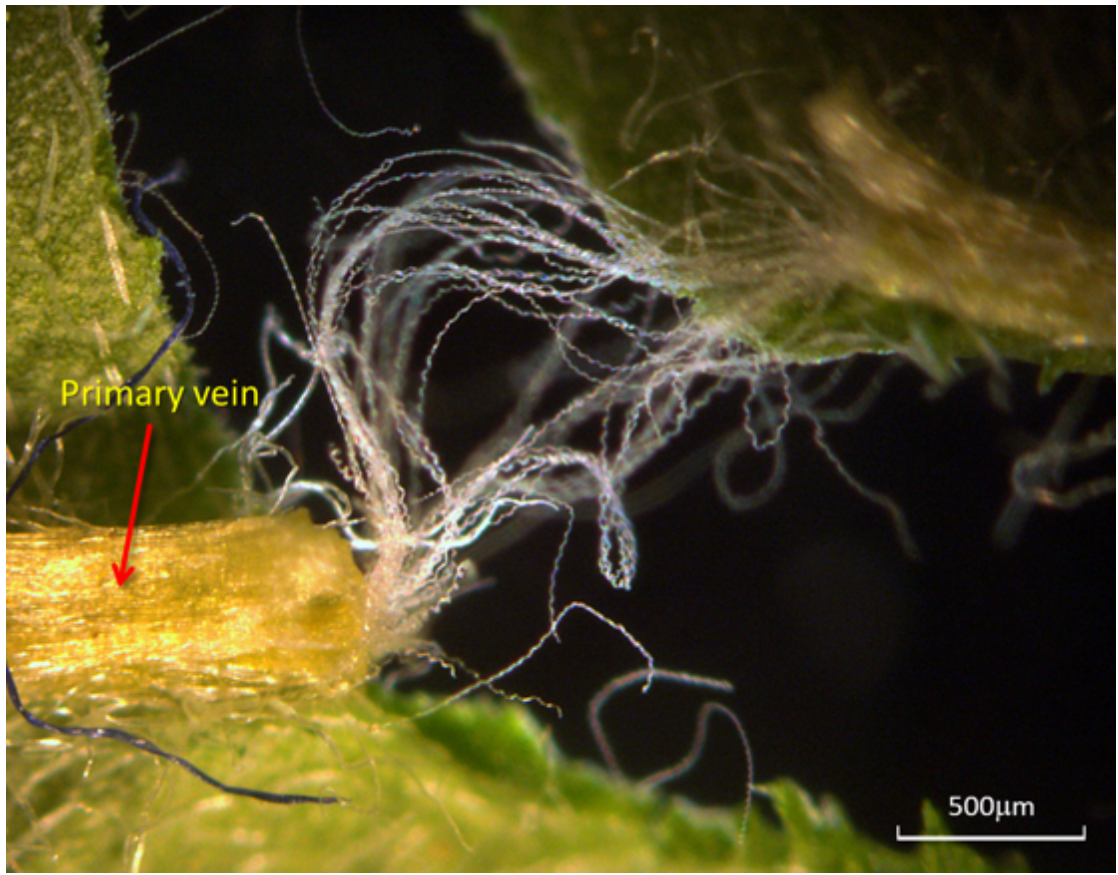

Figure 13. Microscopic image of the spiral fibers connecting two sides of the fractured secondary vein, still intact after high axial strains. [Images by Mehrashk Meidani, SEM Center at USC, Columbia].

## 6   Proposed testing method

Leaves are living organisms; their nonlinear viscoelastic-viscoplastic mechanical properties, including the fracture toughness, may change by time and are loading-rate dependent. Moreover, since the leaf structure is composed of several structural elements with different sizes, the target area of the leaf should be defined for testing. The best approach would be to take several specimens with different sizes and define the fracture toughness as a function of the specimen size. However, this approach is not favorable from an engineering point of view; engineers like to deal with size-independent parameters in their analyses and designs. This leads to the idea that punch-and-die test method could not be an appropriate testing method for highly nonhomogeneous materials like leaves. Yet, there are leaves that are homogeneous at intermediate scale, and the punch-and-die method could be useful in finding their fracture toughness. The following considerations should be taken into account when developing a test method for evaluating the fracture toughness of leaves:

- Leaves should be stored in a temperature and humidity control chamber after picking from the plant and before preparing the specimen. Loss of moisture causes the fracture toughness of the tissues to change substantially [Vincent, 1983].
- The time between testing and picking the leaf from the plant should be standardized since different chemical reactions start to occur right after detaching the leaf from the mother plant.
- The range of force needed to grow a crack in a leaf could vary two orders of magnitudes, similar to the most of other living tissues having a composite structure.
- Rate dependency of the leaf fracture toughness should be studied and speed of testing has to be standardized.
- For a punch-and-die setup, several punch head diameters are needed to test the leaf fracture toughness at several structural scales.
- Clamping parts of the device should not destroy the tissues of the leaf, because it affects the fracture properties of the specimen [Aranwela et al. 1999].

## 6.1 Application of digital image correlation (DIC) technique in fracture testing of leaves

Digital image correlation (DIC) is a non-invasive measurement technique that gives a full three dimensional field of strain on the test specimen [Sutton et al. 1986, 1991] . This technique has been used extensively in the past two decades for fracture testing of different materials including metals [Sutton et al. 2007] and biomaterials (e.g. blood veins).  In order to get reasonable results from DIC method, the subject should have a high contrast pattern on its surface of study [Cheng et al. 2002]. One can spray non-petrol based paint on the leaf surface with two different contrasting colors. While speckling the leaf surface in this way may increase the resolution of data, it can affect the mechanical properties of the leaf if painting process alters properties of leaf tissues chemically or physically, especially during the drying period of paint. More study is required on the effects of speckling process with various chemicals on the mechanical properties of leaves. For now, the author recommends a simple marking method and assume that in such a short time before performing the test, marker chemicals cannot react with leaf tissues and alter its mechanical properties. Just like what is done in crack opening displacement method, we can track several marked points on two sides of the crack line. However, since the crack growth path is not well predictable ahead of the test, several points should be marked on the leaf surface. Using a high speed camera, the behavior of the leaf during the fracture process can be recorded and the data can be used to verify simulation results.

## 7  Concluding remarks

Leaves are complicated composite structures. In order to study the fracture mechanism in leaves, one should know the mechanical properties of different structural elements of the leaf, including veins with various diameters and the intercoastal (membrane) material. Each of these elements behaves in its own way and the global fracture behavior of a leaf is a nonlinear combination of the fracture behavior of its structural elements. In this study, the author briefly described the leaf structure from a mechanical perspective and then identified the parameters that may affect its fracture toughness. Two conventional testing methods used to evaluate the fracture toughness were reviewed and suggestions were given on the ways those methods could be improved.

The qualitative results of a simple fracture experiment on a sample leaf were presented along with the observed macroscopic and microscopic fracture growth mechanisms. Finally, microscopic DIC technique was suggested for quantitative studying of the fracture growth mechanism in the leaves.

# Bibliography

Some of the references are only stated in the captions of the figures since they were obtained from various internet sources. The rest are listed below: